\documentclass[%
 aip,
 amsmath,amssymb,
reprint,%
]{revtex4-1}

\usepackage{graphicx}
\usepackage{dcolumn}
\usepackage{comment}
\usepackage[utf8]{inputenc}
\usepackage[T1]{fontenc}
\usepackage{mathptmx}
\usepackage{etoolbox}
\usepackage{subcaption}
\usepackage{siunitx}
\usepackage[colorlinks,linkcolor=dblue,citecolor=dgreen]{hyperref}
\usepackage[normalem]{ulem}

\usepackage[dvipsnames,svgnames,table,hyperref]{xcolor} 
\definecolor{dblue}{rgb}{0 0.447 0.741}
\definecolor{dgreen}{rgb}{0.274 0.784 0.424}
\definecolor{dred}{rgb}{0.7 0 0.1}

\makeatletter
\def\@email#1#2{%
 \endgroup
 \patchcmd{\titleblock@produce}
  {\frontmatter@RRAPformat}
  {\frontmatter@RRAPformat{\produce@RRAP{*#1\href{mailto:#2}{#2}}}\frontmatter@RRAPformat}
  {}{}
}%
\makeatother
\begin{document}

\preprint{AIP/123-QED}

\title{Helicity oscillations in Rayleigh-B\'enard convection of liquid metal in a cell with aspect ratio 0.5}

\author{Rahul Mitra}
  \affiliation{Helmholtz-Zentrum Dresden-Rossendorf, Bautzner
	Landstr. 400, 01328 Dresden, Germany}
\author{Frank Stefani}
  \affiliation{Helmholtz-Zentrum Dresden-Rossendorf, Bautzner
	Landstr. 400, 01328 Dresden, Germany}
   \email{f.stefani@hzdr.de}
\author{Vladimir Galindo}
  \affiliation{Helmholtz-Zentrum Dresden-Rossendorf, Bautzner
	Landstr. 400, 01328 Dresden, Germany}	
\author{Sven Eckert}
  \affiliation{Helmholtz-Zentrum Dresden-Rossendorf, Bautzner
	Landstr. 400, 01328 Dresden, Germany}
\author{Max Sieger}
  \affiliation{Helmholtz-Zentrum Dresden-Rossendorf, Bautzner
	Landstr. 400, 01328 Dresden, Germany}
\author{Tobias Vogt}
  \affiliation{Helmholtz-Zentrum Dresden-Rossendorf, Bautzner
	Landstr. 400, 01328 Dresden, Germany}
\author{Thomas Wondrak}
  \affiliation{Helmholtz-Zentrum Dresden-Rossendorf, Bautzner
	Landstr. 400, 01328 Dresden, Germany}
%

\begin{abstract}
In this paper, we present numerical and 
experimental results on helicity oscillations 
in a liquid-metal 
Rayleigh-B\'enard (RB) convection cell, 
with an aspect ratio of \num{0.5}. 
We find that helicity oscillations occur 
during transitions of flow states 
that are characterised by significant 
changes in the Reynolds number.
Moreover, we also observe helicity oscillations 
at flow conditions 
where the temporal gradient of 
the change in the Reynolds 
number is significantly smaller 
than that of the helicity.
Notably, the helicity oscillations 
observed during the transient 
double-roll state exhibit characteristics 
remarkably similar to those 
associated with the Tayler Instability. 
\end{abstract}
\maketitle
\section{Introduction}
Rayleigh-B\'enard (RB) convection is a key paradigm of fluid dynamics.
A fluid volume 
(with height $\mathit{H}$ and diameter $\mathit{D}$) that is 
heated at the bottom and cooled at the top acquires a temperature 
difference that leads to a motion of the fluid. 
Depending on the aspect ratio ({${\Gamma}$ = $\mathit{D}/\mathit{H}$}), 
this motion can, for instance, take on the form of a flywheel structure with one or multiple 
stacked roll(s) \cite{Cioni1997, Zuerner2019, Zwirner2020} or that of a 
jump rope vortex \cite{Vogt2018,Akashi2019}. 
In either case, it is the scaling of the Nusselt ($\mathit{Nu}$) and 
the Reynolds number ($\mathit{Re}$) with the Rayleigh  ($\mathit{Ra}$) 
and the Prandtl number ($\mathit{Pr}$) of the liquid which is of highest 
relevance.
Grossmann and Lohse \cite{Grossmann2000} had set-up a systematic theory 
for those scalings, distinguishing four regimes (with some subdivisions) in 
the parameter space of $\mathit{Ra}$ and $\mathit{Pr}$, defined by 
whether the boundary layer or the bulk dominates the global kinetic 
and thermal dissipation. 
In this context, liquid metal experiments are particularly suited to 
explore the region of low Prandtl numbers in the universal 
Grossmann-Lohse phase diagram. 
Pertinent experiments were carried out
with mercury \cite{Takeshita1996,Cioni1997,Glazier1999,Tsuji2005}, 
pure gallium \cite{King2013,Vogt2018}, the eutectic alloy
GaInSn \cite{Akashi2019,Zuerner2019,Schindler2022,Schindler2023,Ren2022},
and sodium  \cite{Khalilov2018}.
For aspect ratio {${\Gamma}$ = \num{1}}, the experiments 
by Ren et al. \cite{Ren2022} revealed a scaling of 
$\mathit{Nu}\sim \mathit{Ra}^{\num{0.25}}$, in correspondence 
with the predictions of the Grossmann-Lohse theory for their 
so-called $I_I$ regime.
By contrast, convection in a more slender cylinder 
with {${\Gamma}$ = \num{0.5}} shows a 
modified scaling  $\mathit{Nu}\sim \mathit{Ra}^{\num{0.28}}$ 
\cite{Glazier1999,Schindler2022,Schindler2023}.
In this geometry the resulting turbulent flow is 
characterized 
by chaotic transitions between single, 
double and triple rolls, as was recently revealed by 
employing Contactless Inductive Flow Tomography (CIFT) 
for reconstructing the global 3D flow field 
\cite{Wondrak2018,Wondrak2023}.

A second aspect of liquid metal experiments 
\cite{Cioni2000,Aurnou2001,Burr2001,Zuerner2020,Vogt2021} 
is their intricate connection to magnetohydrodynamics.
In a recent "pub crawl" through the parameter space, 
using the RoMag device 
at the University of California, the interplay between 
convective, magnetic, and rotational forces was 
explored in much detail \cite{Grannan2022,Schumacher2022}.
However, the impact of magnetic fields on convection 
reflects only one direction of magnetohydrodynamic interactions. 
Conversely, convection is also considered an essential 
ingredient of magnetic field generation 
in planets and stars via hydromagnetic dynamo action 
\cite{Rincon2019,Tobias2021}. 

A decisive role in this process is played by the 
helicity of the flow \cite{Moffatt2018}, i.e. the scalar product of the 
velocity and vorticity, which is capable of inducing
electrical currents {\it in the direction} of a prevailing magnetic 
field. 
This so-called $\alpha$-effect represents a key ingredient 
of $\alpha^2$-dynamos, such as the geodynamo, as well as 
of $\alpha-\Omega$-dynamos, such as the solar dynamo.
While for a long time the focus of dynamo theory was
mainly on a constant $\alpha$-effect, the 
interest has recently shifted to the role of its
temporal variations.
For an early treatment  
of an oscillatory $\alpha$-effect in the context of the 
solar dynamo, see the work by Bushby and Proctor \cite{Bushby2010}.

Returning to convection, it is quite obvious 
that the sloshing motion of Large-Scale Circulation (LSC), 
with its oscillating sidewise deflection of the single-roll 
"flywheel", is connected with a corresponding helicity 
oscillation. 
A similar helicity oscillation was also observed 
in simulations of the current-driven, kink-type Tayler Instability 
(TI), first for the simple case of a full, non-rotating
cylinder \cite{Weber2015}, but later also 
for a quite realistic model of a rotating star 
\cite{Monteiro2023}.
Remarkably, in either case the helicity oscillation of the dominant 
flow mode with its azimuthal wavenumber {$m$ = \num{1}} goes along 
with a negligible change of the energy content of the flow. 

This observation has motivated recent modelling efforts to 
explain the amazing synchronicity of the solar 
Schwabe cycle with the \num{11.07}-year alignment cycle of 
the tidally-dominant planets, 
Venus, Earth and Jupiter, in terms of 
an energy-efficient tidal entrainment mechanism for 
the helicity at the solar tachocline \cite{Stefani2016,Stefani2018,Stefani2019,Stefani2020,Stefani2021,Klevs2023,Stefani2023}.
In this context one may ask whether tidal helicity synchronization
can be observed in the laboratory.
In view of the significant technical challenges to perform TI experiments with 
liquid metals \cite{Seilmayer2012}, a first attempt to study this 
effect in the lab was made on the basis of a classical RB-flow 
in a {${\Gamma}$ = \num{1}} cylinder, exposed to a tide-like ({$m$ = \num{2}}) 
force exerted by two oppositely situated coils \cite{Juestel2020,Juestel2022}.
At a critical strength of the "tidal" force, synchronization of 
the helicity of the flow was indeed observed (although in a spatially 
segregated manner).

But what about helicity, and its potential synchronization, in 
RB-flows with {${\Gamma}$ < \num{1}}, which are typically characterized by 
transitions between single, double and triple rolls \cite{Wondrak2023}?
Leaving the synchronization issue to future work,  
this paper is primarily concerned with the time evolution 
of the helicity in {${\Gamma}$ = \num{0.5}} convection, its 
specification for single, double and triple rolls, and the 
transitions between them.

We will start with numerical simulations for {$\mathit{Ra}$ = \num{1.8e7}} 
for which 
we will characterize in detail helicity reversals 
(i.e., half oscillations) for the 
two cases of a single vortex flow and a double vortex flow. 
Informed by that, we will go over to an experiment carried out at the 
much higher Rayleigh number
{$\mathit{Ra}$ = \num{6.02e8}}, for which we will determine the 3D flow
by CIFT.
After shortly recalling the experimental set-up of the 
\mbox{${\Gamma}$ = \num{0.5}} experiment \cite{Wondrak2023}, we 
will analyse exemplary helicity reversals of the experimental flow.
%
\section{Simulation}\label{sec:simulation}
\subsection{Simulation setup}
The flow in a RB cell of height \SI{640}{\milli\metre} and 
diameter \SI{320}{\milli\metre} was directly simulated (without 
any turbulence modelling) for more than \SI{6}{\hour} using 
the pisoFOAM solver of the finite volume library OpenFOAM. 
For the simulations a very fine mesh consisting of \num{1563028} 
cells was used.
In order to cover a long period with a reasonable 
numerical effort and calculation time, 
the temperature difference was limited to 
 \SI{0.2}{\kelvin}, which is equivalent 
to a Rayleigh number {$\mathit{Ra}$ = \num{1.8e7}}.
Subsequently, the very finely discretized OpenFOAM velocities were 
interpolated on a coarser cylindrical mesh, containing \num{5760} 
linear hexahedral elements with a constant edge length of 
\SI{25}{\milli\metre} over the height and a mean edge length 
of \SI{20}{\milli\metre} over the diameter, with a total of 
{$n_{\mathit{vol}}$ = \num{6625}} discretized points. 
These interpolated velocity fields then served for further 
analysis, in particular for comparisons with respective 
CIFT results.
They were stored at every second, and 
for our analysis we considered the 
time period from 
\SI{8000}{\second} to \SI{21600}{\second}
when the convective flow was safely established.
\subsection{General flow characteristics}

The typical time scale to characterize the dynamics of 
an LSC is the free-fall time $t_{\mathit{ff}}$, i.e. 
the time taken for a hot plume to rise from bottom to 
top according to the density difference which, in turn, depends 
on the temperature difference:
\begin{equation}\label{eqn:freefall_time}
	t_{\mathit{ff}} = \sqrt{\frac{\mathit{H}}{g \alpha \Delta T}}.
\end{equation}
Here, $g$ is the acceleration due to gravity, $\mathit{H}$ is the height 
of the cell, $\alpha$ is the coefficient of volume expansion, 
and $\Delta T$ is the temperature difference between top and 
bottom plate. 
For our specific simulation, the free-fall time 
is \SI{51.4}{\second}.

The global, volume averaged Reynolds number of the flow was calculated 
using the following expression:
\begin{equation}
	\mathit{Re}_{\mathit{vol}} = \frac{\bar{\mathit{v}} \mathit{H}}{\nu},
\end{equation}
where $\bar{\mathit{v}}$ is the volume averaged flow velocity, 
and $\nu$ is the kinematic viscosity.

The mean helicity density of the flow was calculated using the 
following expression:
\begin{equation}
	h = \frac{\sum_{i=1}^{n_{\mathit{vol}}} w_i(\mathbf{v}_i \cdot (\nabla \times \mathbf{v}_i))}{\sum_{i=1}^{n_{\mathit{vol}}} w_i},
\end{equation}
where $w_i$ is the volume associated with each element of the 
coarser cylindrical grid, $\mathbf{v}_i$ is the velocity 
in this element $i$, and $\nabla \times \mathbf{v}_i$ is the vorticity 
therein. 
As a density of helicity, $h$ has the dimension 
 [$LT^{-2}$] and will always be given in units mm s$^2$.
Henceforth, we will frequently use the term 
"helicity" for the "mean helicity density",
although the former would, strictly speaking, be the volume integral
over the local helicity density.

In order to get a preliminary overview about the
flow modes that typically appear in our problem,
we have carried out a proper orthogonal decomposition 
(POD), similar to the formulation described by Wondrak et al. \cite{Wondrak2023}. 
\begin{figure}[h!]
    \includegraphics[width=1\linewidth]{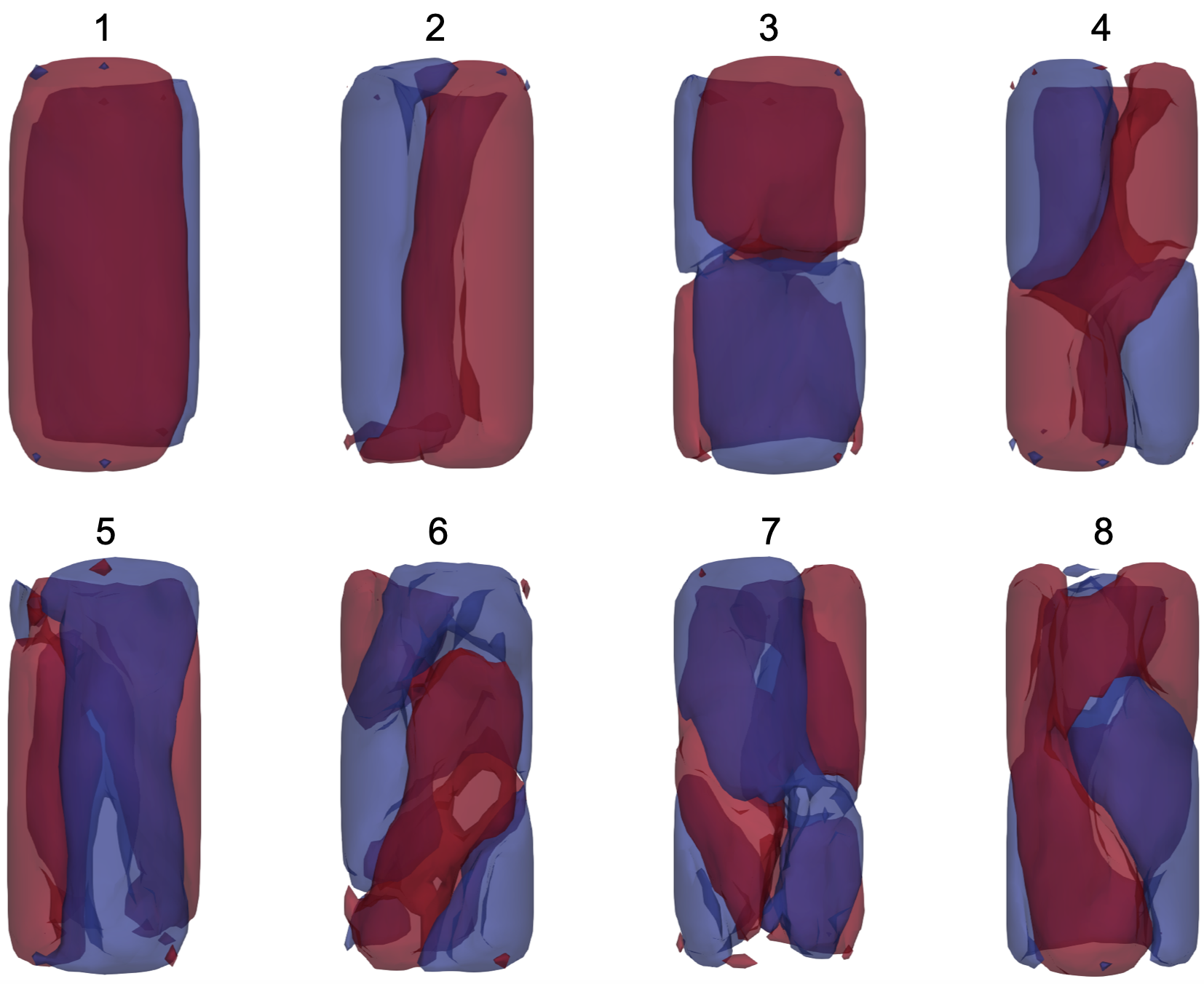}
    \caption{The first \num{8} POD modes extracted from the simulated velocity field: 
    3D isosurfaces of the vertical velocity referred to a value corresponding to 
    $\pm$\num{0.1} of the respective maximum value.}
    \label{fig:pod_modes_sim}
\end{figure}
The first \num{8} POD modes are shown in Fig. \ref{fig:pod_modes_sim}.
Modes \num{1} and \num{2} resemble a single roll state (SRS), while 
modes \num{3} and \num{4} resemble a double roll state (DRS).
Mode \num{6}, in turn, is a twisted structure, dominated by 
three rolls over 
the height, 
which will therefore be called triple roll state (TRS).
The rest of the modes \num{5}, \num{7} and \num{8} represent other
structures which cannot easily be classified as a SRS, DRS or a TRS.
Modes above \num{8} 
contribute to less than 
\num{2}\% of the total energy.
The individual energy contributions of the SRS, DRS and TRS were 
calculated over time.
%
\subsection{Time evolution}
\begin{figure}[h!]
   \includegraphics[width=1\linewidth]{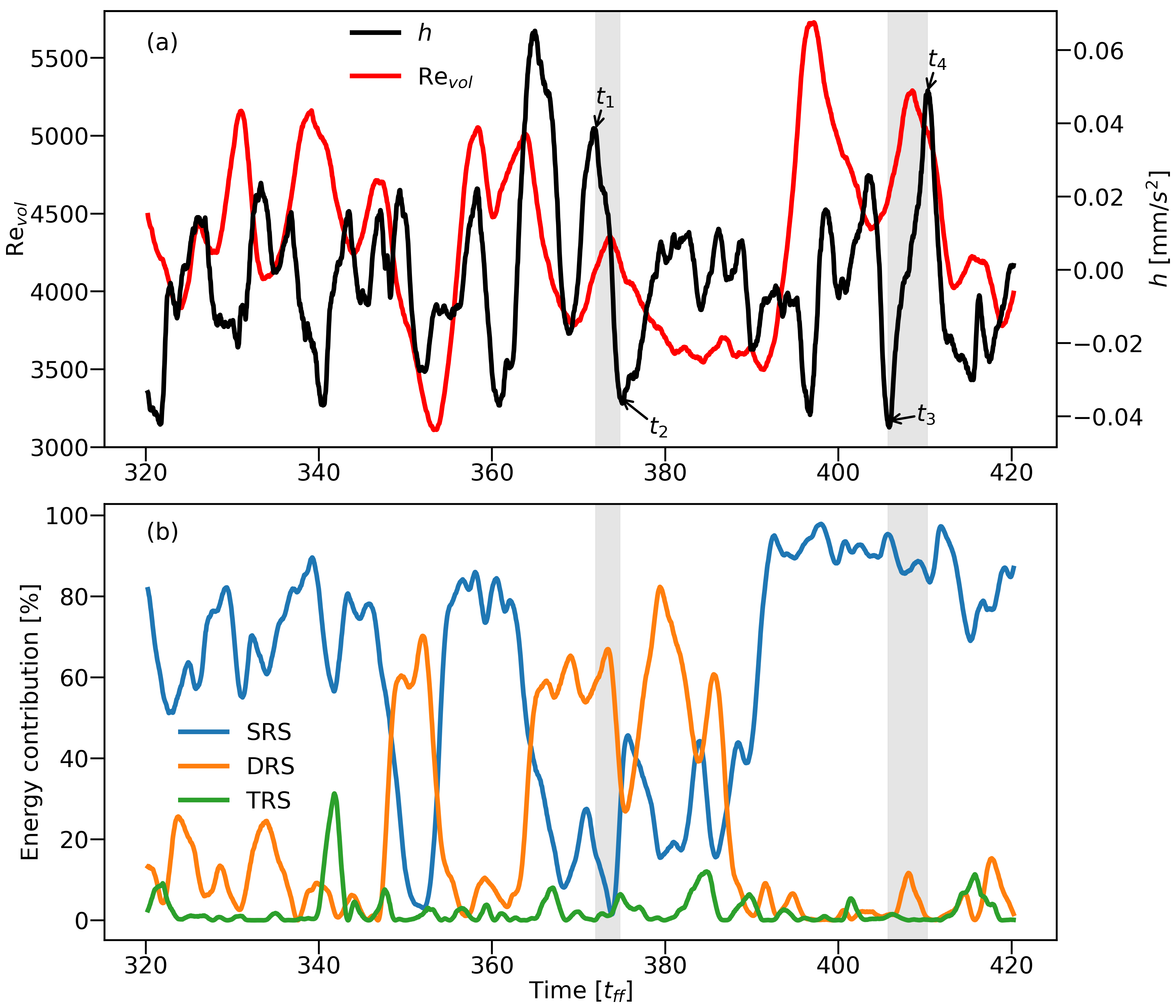}
    \caption{(a) Global Reynolds number (red) and mean helicity density (black) 
    over time of the simulated velocity field. 
    (b) Energy contributions of SRS (blue), DRS (orange), TRS (green) 
    to the overall flow structure. 
    The time is given in units of $t_{\mathit{ff}}$. 
    Two exemplary helicity reversals are shaded in grey, 
    between $t_1$ - $t_2$ and $t_3$ - $t_4$.}
    \label{fig:h_re_modes_sim}
\end{figure}

The global Reynolds number $\mathit{Re}_{\mathit{vol}}$ and the 
mean helicity density $h$
over a representative time interval of 
\num{100} free-fall times from the 
entire simulation are shown in Fig. \ref{fig:h_re_modes_sim}(a).
While the Reynolds number fluctuates only moderately between values 
of \num{3100} and \num{5600},
the mean helicity density fluctuates strongly, acquiring both 
positive and negative values 
with amplitudes of up to \SI{0.06}{\milli\metre\per\square\second}.
Quite often one observes
concurrent changes of Reynolds number and helicity which, on 
closer inspection, correspond to transitions 
between flow modes with different roll numbers.

However, there are also periods where the helicity 
fluctuates significantly, 
while the flow structure remains relatively stable
and the rate of change of the Reynolds number 
is quite small.
Two such time intervals are marked in Fig. \ref{fig:h_re_modes_sim}(a), 
where between $t_1$ - $t_2$ ($\approx$ \num{2.8} $t_{\mathit{ff}}$) 
and $t_3$ - $t_4$ ($\approx$ \num{4.6} $t_{\mathit{ff}}$) the mean 
helicity density 
changes its sign, 
while the Reynolds number is quite stable.
Before analyzing those intervals in more detail,  
we show in Fig. \ref{fig:h_re_modes_sim}(b)
the energy contributions of the individual 
flow modes over time.

Quite generally, the SRS, represented in blue, is the most dominant mode, 
often reaching more than \num{80}$\%$ of the total energy. 
The pattern of fluctuation does not appear to be periodic, 
being an indication for, in general, chaotic dynamics of the flow.
The DRS, shown in orange, also exhibits fluctuations, but with a generally 
lower percentage of the total energy compared to the SRS. 
The energy level of the DRS fluctuates 
between \num{20}$\%$ and \num{60}$\%$. 
The TRS, depicted in green, contributes the least energy among the 
three states, with most values remaining below \num{20}$\%$. 
The strong changes in the energy shares occurring at certain times 
are a symptom of the transformation between the different flow states.
In Appendix \ref{appendix:fourier_simulation}, we will discuss 
the dominant frequencies of those variations.
%
\subsection{Helicity reversals}
Helicity in fluid dynamics refers to the "handedness'', or chirality, 
of the flow structure, which can be understood as the direction in 
which the flow twists or spirals. 
A reversal of helicity implies that a flow structure that was twisting in
one direction is now twisting in the opposite direction.

\begin{figure}[h!]
    \includegraphics[width=1\linewidth]{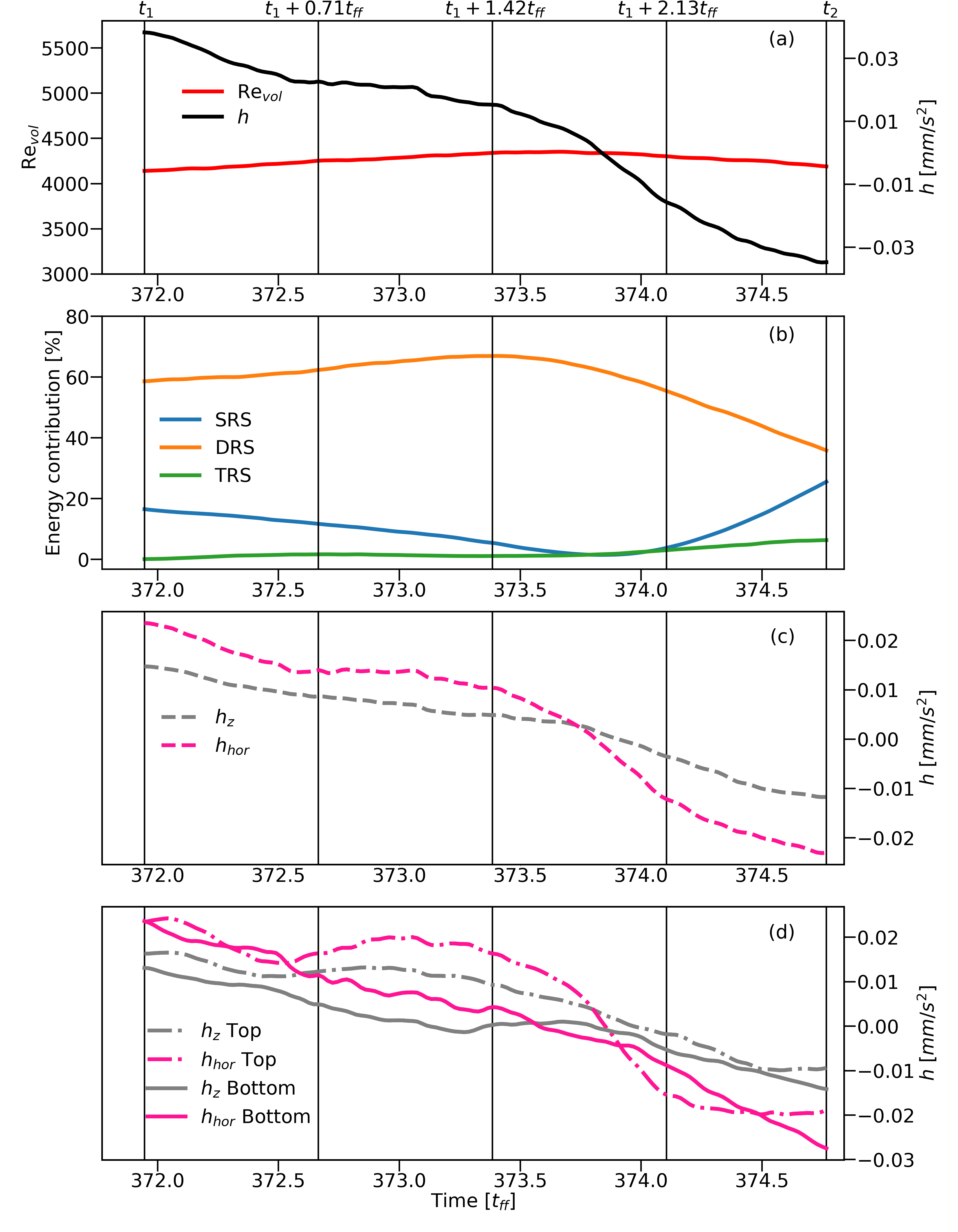}
    \caption{(a) Global Reynolds number (red) and mean helicity density (black); 
    (b) Energy contributions of SRS (blue), DRS (orange), TRS (green); 
    (c) Vertical (dashed grey) and horizontal components of the mean helicity density (dashed pink); 
    (d) Vertical (grey) and horizontal (pink) components of the mean helicity density at 
    top half of the cylinder (dash-dotted) and bottom half of the cylinder (solid), 
    within the time interval $t_1$ to $t_2$ 
    of the simulated flow.}
    \label{fig:sim_hr_hz_t1_t2}
\end{figure}

In the following, we will discuss those helicity reversals 
(i.e., half oscillations) that take place 
within a time interval when the 
dominant type of flow mode mainly persists.
Since we have not observed any long enough interval with 
a dominant TRS, we will focus exclusively on the SRS and DRS.
We start with the latter, for which we 
detail in Fig. \ref{fig:sim_hr_hz_t1_t2} the helicity evolution  
between $t_1$ - $t_2$ (cp. Fig. \ref{fig:h_re_modes_sim}(a)).
As it is obvious from Fig. \ref{fig:sim_hr_hz_t1_t2}(a), 
the mean helicity density reversal comes about
{\it without} any significant change in the Reynolds number 
(just around \num{5}\%) which 
indicates that the flow maintains its overall energy.
It is also seen in 
Fig. \ref{fig:sim_hr_hz_t1_t2}(b) 
that the DRS is dominant until
the end of the interval when 
the SRS starts to gain comparable strength.

Similarly as in the work by 
J\"ustel et al.\cite{Juestel2022}, 
we distinguish here between 
the vertical and the horizontal contributions of the mean 
helicity density, $h_z$ and $h_{hor}$, 
which result from $v_z  \cdot (\nabla \times \mathbf{v})_z$
and $v_x \cdot (\nabla \times \mathbf{v})_x+v_y \cdot (\nabla \times \mathbf{v})_y$, 
respectively. 
In our problem, $h_{hor}$ represents mainly the sloshing motion
of the LSC, characterized by a side-wise deflection of the large-scale roll 
with its dominant horizontal vorticity. $h_z$, in turn, corresponds to 
some torsional motion where the large-scale roll is tilted and thereby 
acquires a vertical vorticity component $(\nabla \times \mathbf{v})_z$
(some corresponding 
vertical motion is then still required to produce a 
non-zero $h_z$).

As can be seen in Fig. \ref{fig:sim_hr_hz_t1_t2}(c),
these two contributions evolve similarly, although the amplitude of
$h_{hor}$ is about two times larger than $h_z$.
Still more details can be observed in Fig. \ref{fig:sim_hr_hz_t1_t2}(d),
where we further distinguish between the values of  $h_{hor}$ and $h_z$ 
in the top and in the bottom half of the cylinder. 
Basically, $h_z$ and $h_{hor}$ show a quite comparable behaviour 
in the two halves, meaning
that there is no significant cancellation going on.
\begin{figure}[h!]
    \includegraphics[width=1\linewidth]{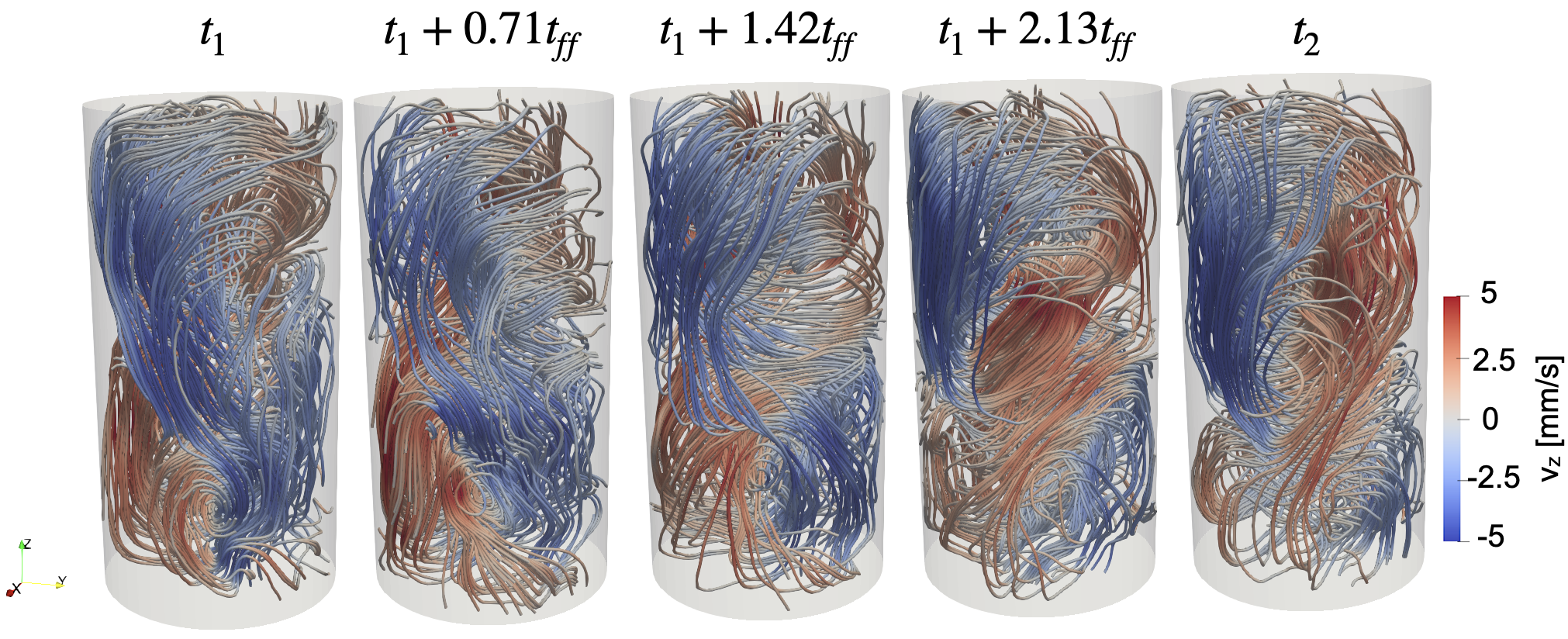}
    \caption{Illustration of the helicity reversal at nearly constant Reynolds number for 
    a DRS, within the time interval $t_1$ to $t_2$ of the simulated flow. The snapshots correspond 
    to the five instants indicated by vertical lines in Fig. \ref{fig:sim_hr_hz_t1_t2}. }
    \label{fig:sim_t1_t2}
\end{figure}
\begin{figure}[h!]
   \includegraphics[width=1\linewidth]{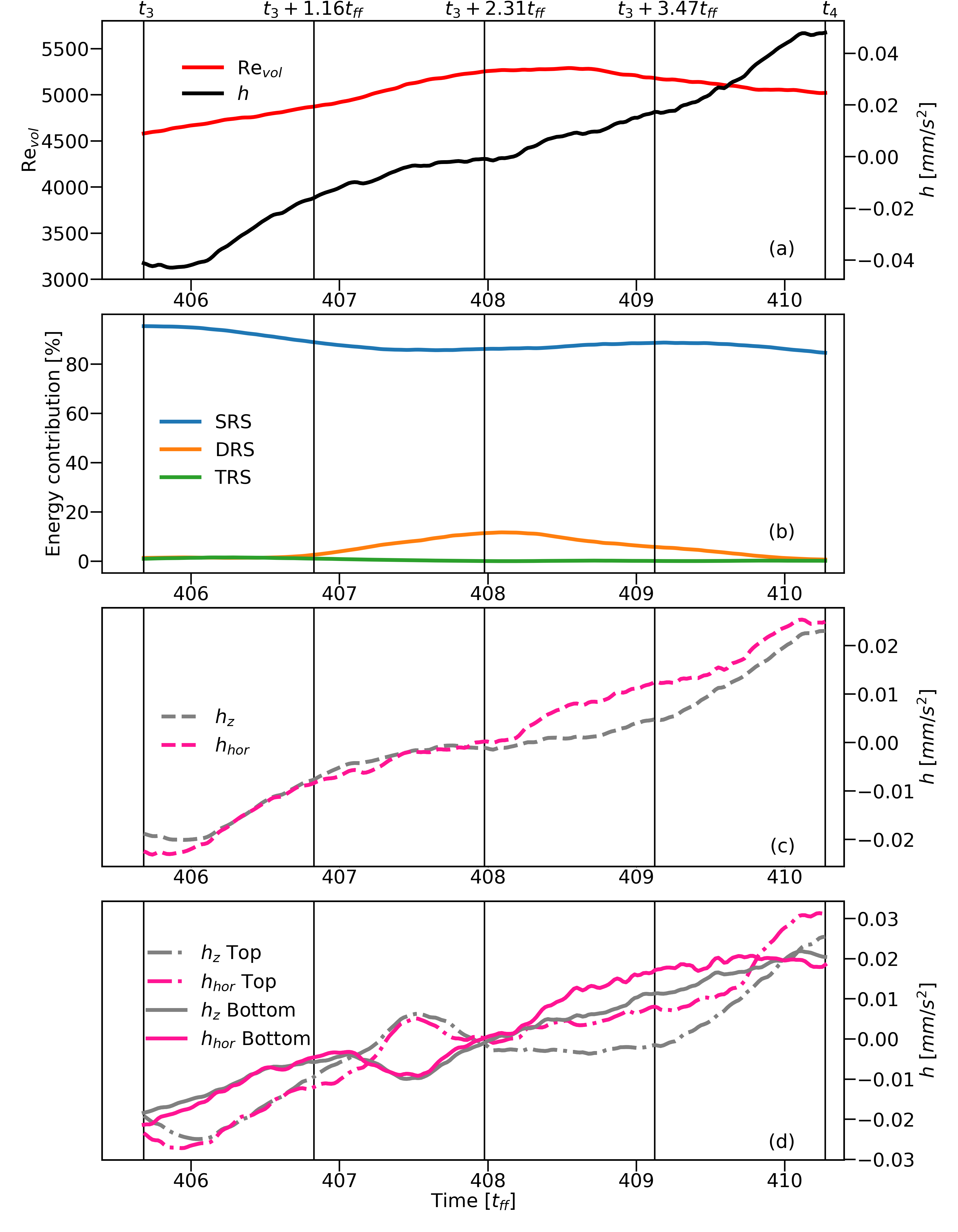}
    \caption{(a) Global Reynolds number (red) and mean helicity density (black); 
    (b) Energy contributions of SRS (blue), DRS (orange), TRS (green); 
    (c) Vertical (dashed grey) and horizontal components of the mean helicity density (dashed pink); 
    (d) Vertical (grey) and horizontal (pink) components of the mean helicity density at 
    top half of the cylinder (dash-dotted) and bottom half of the cylinder (solid), 
    within the time interval $t_3$ to $t_4$ 
    of the simulated flow.}
    \label{fig:sim_hr_hz_t3_t4}
\end{figure}
\begin{figure}[h!]
    \includegraphics[width=1\linewidth]{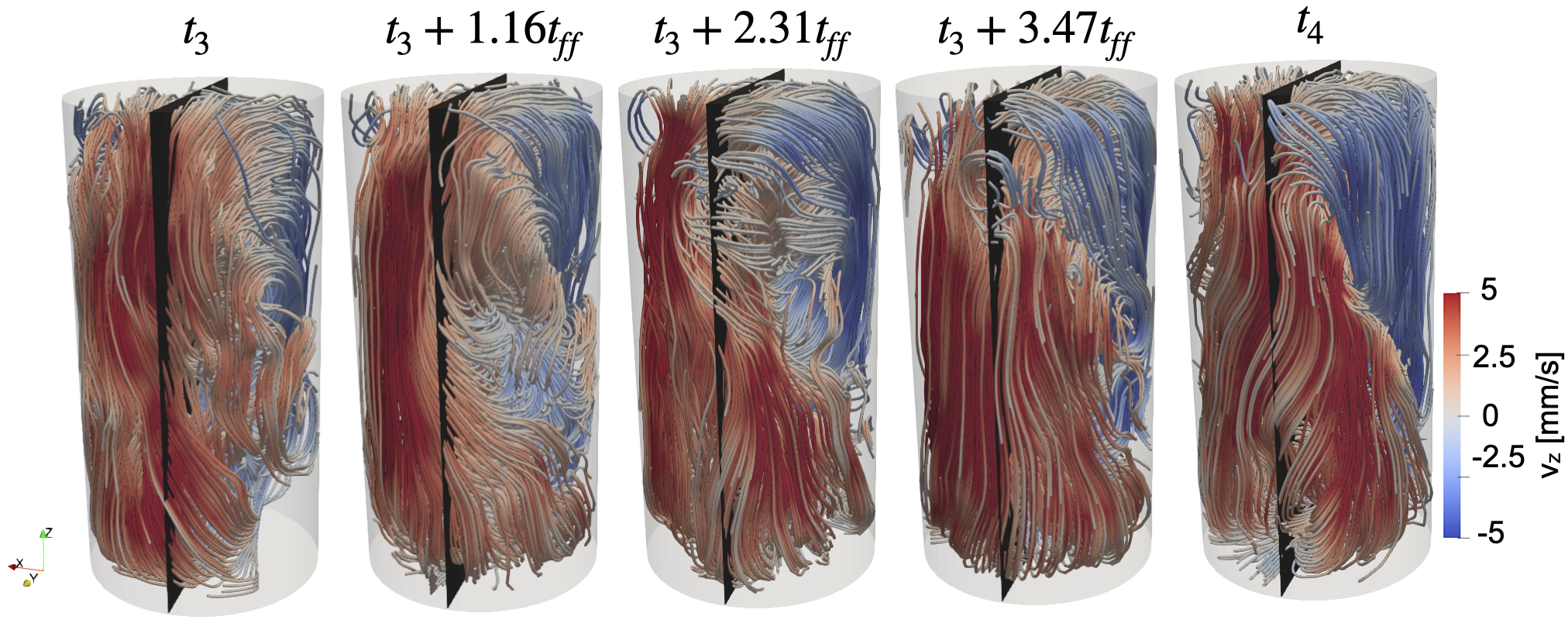}
    \caption{Illustration of the helicity reversal at nearly constant Reynolds number for 
    a SRS, within the time interval $t_3$ to $t_4$ of the simulated flow. The snapshots correspond 
    to the five instants indicated by vertical lines in Fig. \ref{fig:sim_hr_hz_t3_t4}. }
    \label{fig:sim_t3_t4}
\end{figure}
The streamlines in Fig. \ref{fig:sim_t1_t2}, color-coded by the vertical velocity component, 
show the direction of flow and regions of upward (red) and downward (blue) motion
for the five particular instants indicated by vertical lines in Fig.
\ref{fig:sim_hr_hz_t1_t2}. 
Let us start with the flow structure at $t_1$.
Near mid-height of the cell, the red streamlines, indicative of 
upward flow, are positioned behind the blue streamlines, which represent 
a downward flow. 
%
%
As time advances, the red and blue streamlines change their 
relative positions. 
By $t_2$, the red streamlines near mid-height 
have moved to the front, a reversal 
from their original position at $t_1$, 
which indicates that the flow now has an 
opposite helicity compared to the initial state.
The helicity reversal of the DRS 
resembles the 
chirality reversal reported by 
Weber et al. \cite{Weber2015} 
for the case of the kink-type, current-driven TI.
However, the flow structure at $t_2$ is not perfectly mirror-symmetric 
to that at $t_1$, since the SRS has already gained some comparable strength.

The helicity reversal for an SRS, documented in Figs. \ref{fig:sim_hr_hz_t3_t4} 
and \ref{fig:sim_t3_t4}, is a bit different from the case of the 
twisted DRS.
As seen in Fig. \ref{fig:sim_hr_hz_t3_t4}(c),
the two contributions $h_{hor}$ and $h_z$ evolve quite in parallel with
nearly identical contributions. 
The same applies to their shares in the two half-spaces (Fig. \ref{fig:sim_hr_hz_t3_t4}(d)), 
apart from an interesting detail between \num{407} and \num{408} $t_{\mathit{ff}}$ where
the helicities in the two half spaces seem to evolve contrary, thereby 
compensating each other to some extend.

Fig. \ref{fig:sim_t3_t4} shows that 
in case of an SRS, the large-scale "flywheel" undergoes sloshing from 
left to right, whereby the main hot and cold plumes move horizontally over time.
It is also important to note that the time taken for the 
helicity reversal in case of the SRS is longer than the respective 
reversal for twisted DRS.
This is in agreement with the Fourier spectral analysis of the 
helicity and the energy contributions of the SRS and DRS, where the high 
frequency peaks of helicity are dominated relatively more by the DRS as 
compared to the SRS (see Appendix \ref{appendix:fourier_simulation}).
%
\section{Experiment}\label{sec:experiment}
\subsection{Experimental setup}\label{sec:experiment:setup}

\begin{figure}[h]
    \includegraphics[width=1\linewidth]{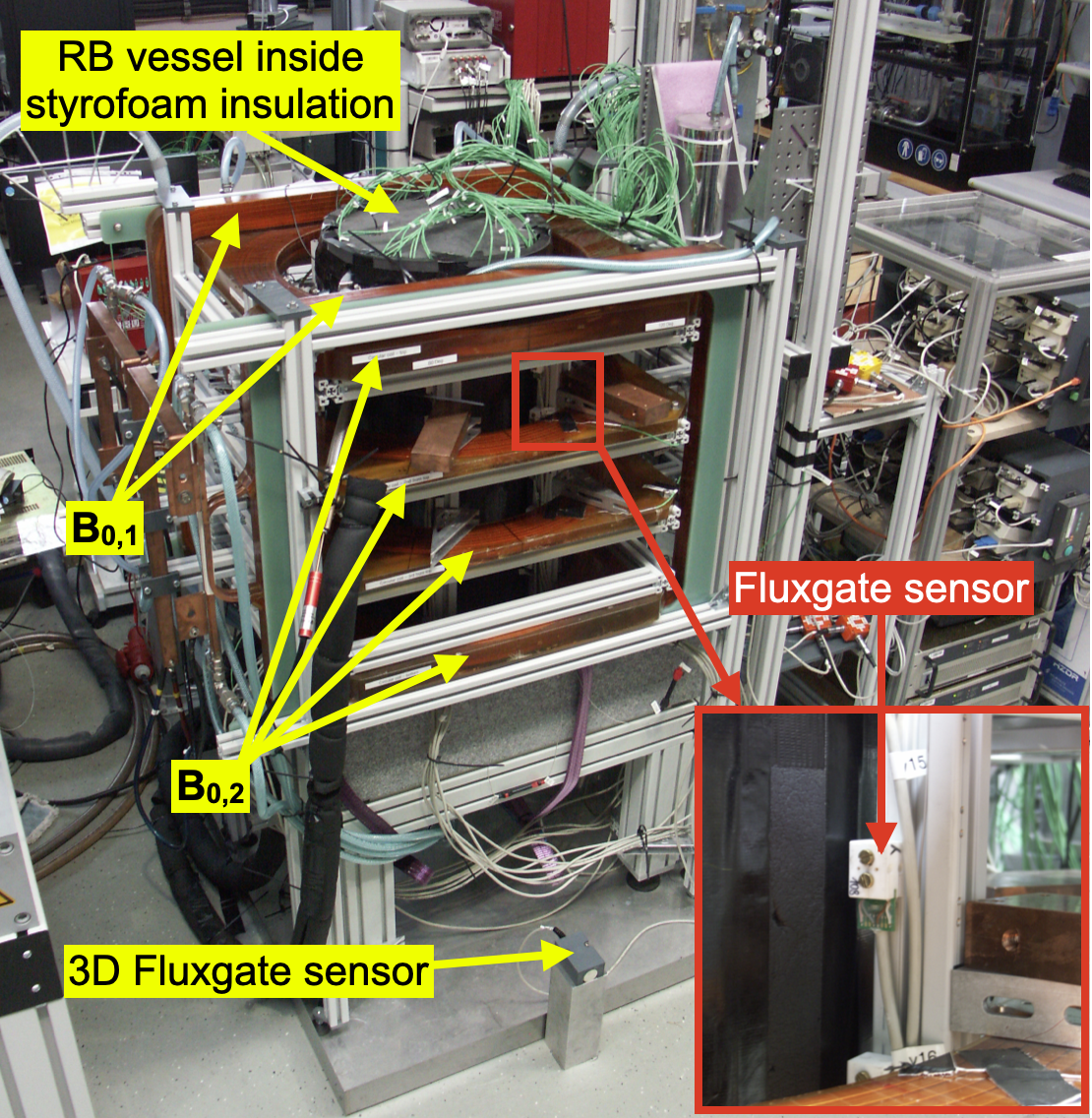}
    \centering
    \caption{The RB experimental system with the CIFT measurement system \cite{Mitra2024}.}
    \label{fig:RB_CIFT}
\end{figure}

This section begins with a short sketch of the experimental set-up, more details 
of which can be found in the work of Wondrak et al. \cite{Wondrak2023}
The RB experimental system is shown in Fig. \ref{fig:RB_CIFT}. 
The cell with a height of \SI{640}{\milli\metre} and a diameter 
of \SI{320}{\milli\metre} is filled with the ternary alloy GaInSn, 
a liquid metal with an eutectic temperature of \SI{10.5}{\celsius}.
In the experiment, the Prandtl number is  
$\mathit{Pr}$ = \num{0.031}.
In order to ensure adiabatic boundary conditions in radial direction, 
the cell is encapsulated 
in styrofoam insulation. 
It has two copper heat exchangers, one at the top and 
the other at the bottom, to achieve 
nearly constant 
temperature boundary conditions
in axial direction.
For the realization of the CIFT measurements, 
four circular excitation coils generate a magnetic field in the 
axial direction, and two rectangular coils generate a magnetic field 
in the horizontal direction.
However, for our experiment, we followed the \emph{one-field} \emph{excitation} 
scheme, described by Mitra et al. \cite{Mitra2024}, 
utilizing only the axial field, which is largely sufficient
to reconstruct the dominant flow structures.
The radial components of the induced 
magnetic fields were 
recorded by \num{42} fluxgate 
sensors arranged in a \num{7}$\times$\num{6} 
(height $\times$ azimuth) configuration 
around the side walls of the cell.
These sets of magnetic fields measured at every 
second were used to reconstruct the 3D velocity at the 
corresponding time instances by solving a linear inverse 
problem \cite{Stefani2000a, Stefani2000b, Stefani2004, Wondrak2018, Mitra2022}, 
using the fast reconstruction algorithm 
to speed up the regularization procedure
\cite{Glavinic2022b}.

The total duration of the experiment was \num{30000} seconds, 
with \SI{1}{\hour} duration at the beginning when the temperature difference 
was zero, followed by \SI{6}{\hour} of flow driven by a temperature difference 
of nearly \SI{12}{\kelvin}, corresponding to a Rayleigh number of 
$\mathit{Ra}$ = \num{6.02e8}, and then again the temperature difference was 
brought down to zero, quickly terminating the flow.

\subsection{General flow characteristics}

The free-fall time for the experimental 
flow, using the definition from Eqn. \ref{eqn:freefall_time}, 
is \SI{8.9}{\second}.
As CIFT is known to provide only a  
global picture of the flow while smoothing out fine 
structures, we have examined 
in Appendix \ref{appendix:helicity_simulation_comparison} 
how reliable it is in reconstructing the helicity of the
flow. 
For that purpose we have used the numerically 
simulated flow field from which we generated a synthetic
CIFT reconstruction. 
As shown in Appendix \ref{appendix:helicity_simulation_comparison}, 
CIFT is in general well capable 
of extracting the helicity fluctuations of the original flow, 
though with the typical underestimation of the 
amplitudes due to the smoothing effect.
\begin{figure}[h]
    \includegraphics[width=1\linewidth]{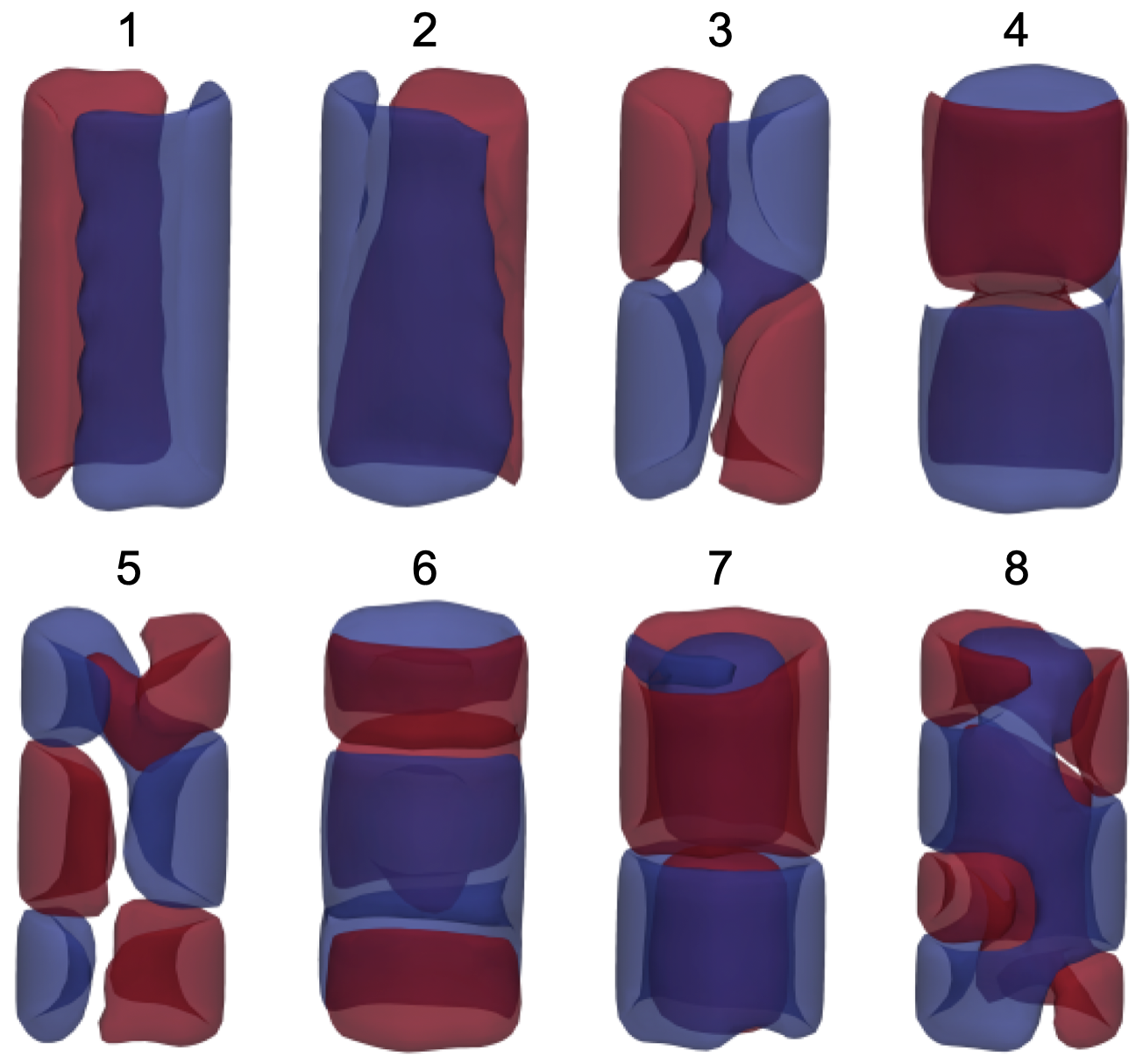}
    \centering
    \caption{The first \num{8} POD modes extracted for the CIFT-reconstructed experimental velocity field \cite{Wondrak2023}: 
    3D isosurfaces of the vertical velocity referred to a value corresponding to 
    $\pm$\num{0.1} of the respective maximum value.}
    \label{fig:pod_modes}
\end{figure}

Similar to the case of simulations, 
the first \num{8} POD modes of the experimental flow, 
as extracted from CIFT, are 
shown in Fig. \ref{fig:pod_modes}. 
Here, modes \num{1} and \num{2} were grouped as SRS, modes \num{3} and \num{4} 
were categorized as DRS and modes \num{5} and \num{6} were classified as TRS.
Modes \num{7} and \num{8} exhibit torus-shaped flow states, 
which have resemblances to a DRS and a TRS, respectively. 
Vertical upward and downward movements 
are still seen on the sidewall, but the recirculation 
of fluid closes in the center zone around the axis 
of the cylinder.
%
\subsection{Time evolution}

\begin{figure}[h]
    \includegraphics[width=1\linewidth]{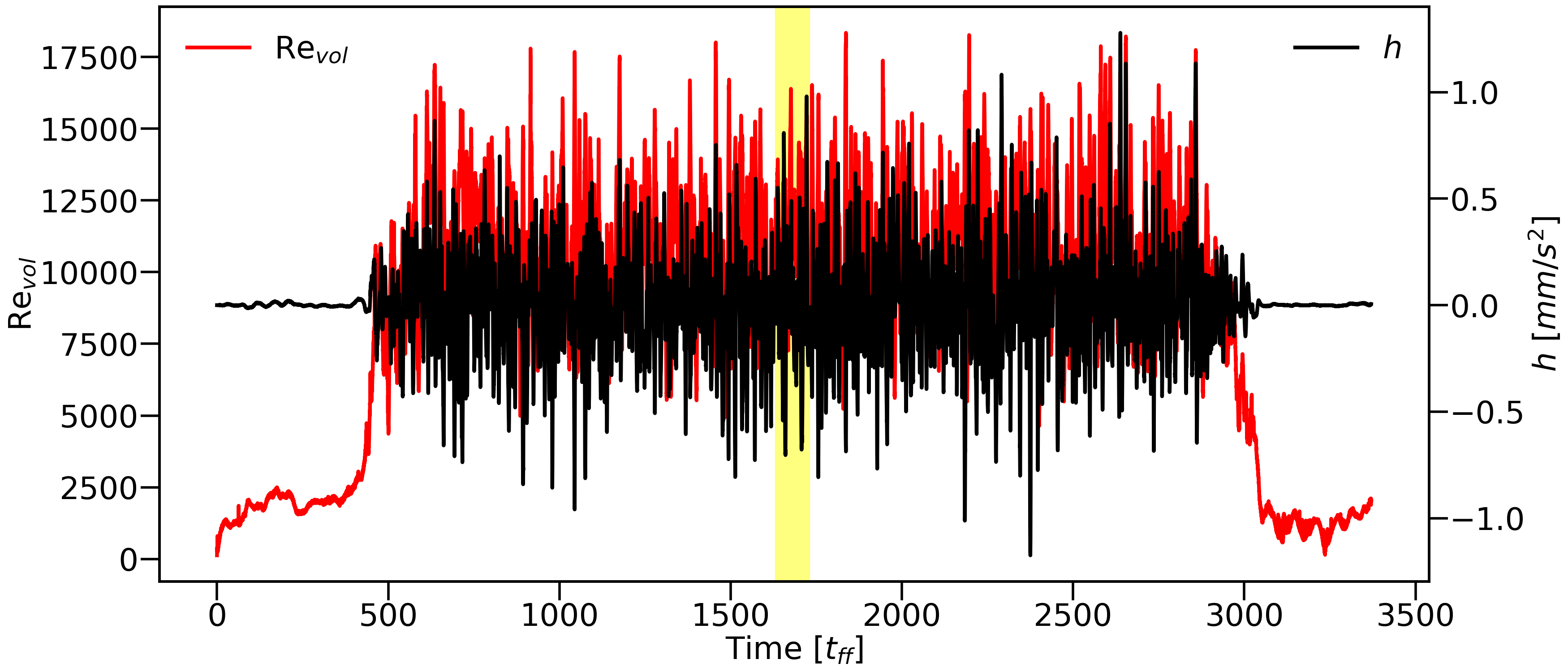}
    \centering
    \caption{Global Reynolds number (red) and mean helicity density
    (black) for the entire duration of the experiment with the \num{100} 
    free-fall duration shaded in yellow.}
    \label{fig:h_re_total}
\end{figure}

The global Reynolds number and the mean helicity density for the 
entire duration of the experiment are shown in Fig. \ref{fig:h_re_total}.
\begin{figure}[h!]
    \includegraphics[width=1\linewidth]{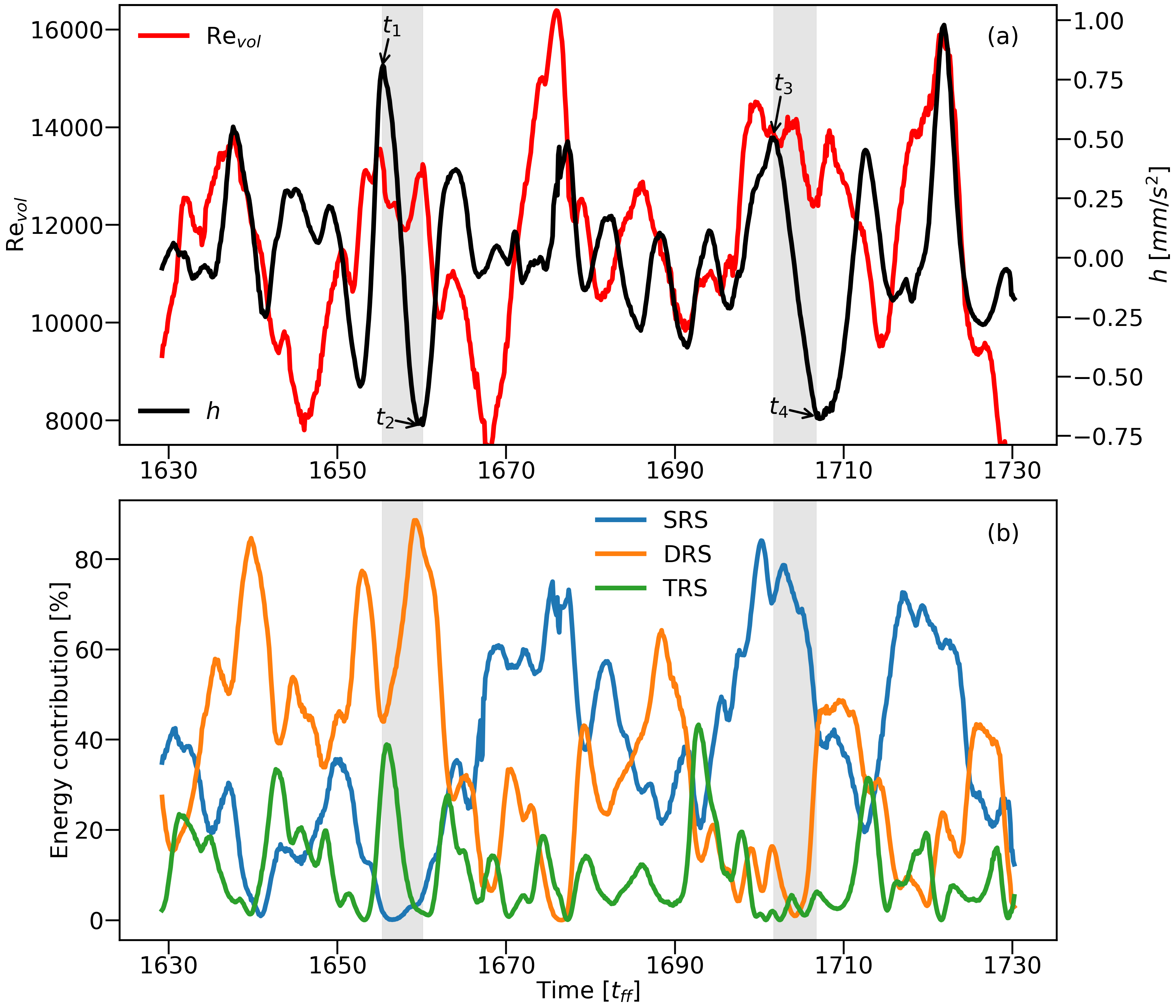}
    \centering
    \caption{(a) Global Reynolds number (red) and mean helicity density (black). 
    (b) Energy contributions of SRS (blue), DRS (orange), TRS (green) 
    to the overall flow structure 
    for \SI{15}{\minute} (\num{100} free-fall time units) 
    of flow (shaded in yellow in Fig. \ref{fig:h_re_total}). 
    Two exemplary helicity reversals are shaded in grey, 
    between $t_1$ - $t_2$ and $t_3$ - $t_4$.}
    \label{fig:h_re_modes}
\end{figure}
During the first hour, the Reynolds number 
(as well as the mean helicity density) of the flow 
remained nearly zero.
Upon setting the temperature difference between top and bottom, 
both mean helicity density and Reynolds number 
began to rise and fluctuate, signaling the onset of fluid motion 
and the development 
of a turbulent flow.

For the restricted time interval, marked by the
yellow area in Fig. \ref{fig:h_re_total}, Fig. \ref{fig:h_re_modes}(a)
shows the global Reynolds number and the helicity over time. 
Correspondingly, in Fig. \ref{fig:h_re_modes}(b) we 
plot the dominant contributions to the flow.
Just as in the simulations discussed above, the three flow states exhibit 
distinct behaviors over time, with the SRS modes being the most energetic and 
dynamic, followed by the DRS modes, and with the TRS modes being the least energetic.
\subsection{Helicity reversals}
\begin{figure}[h!]
    \includegraphics[width=1\linewidth]{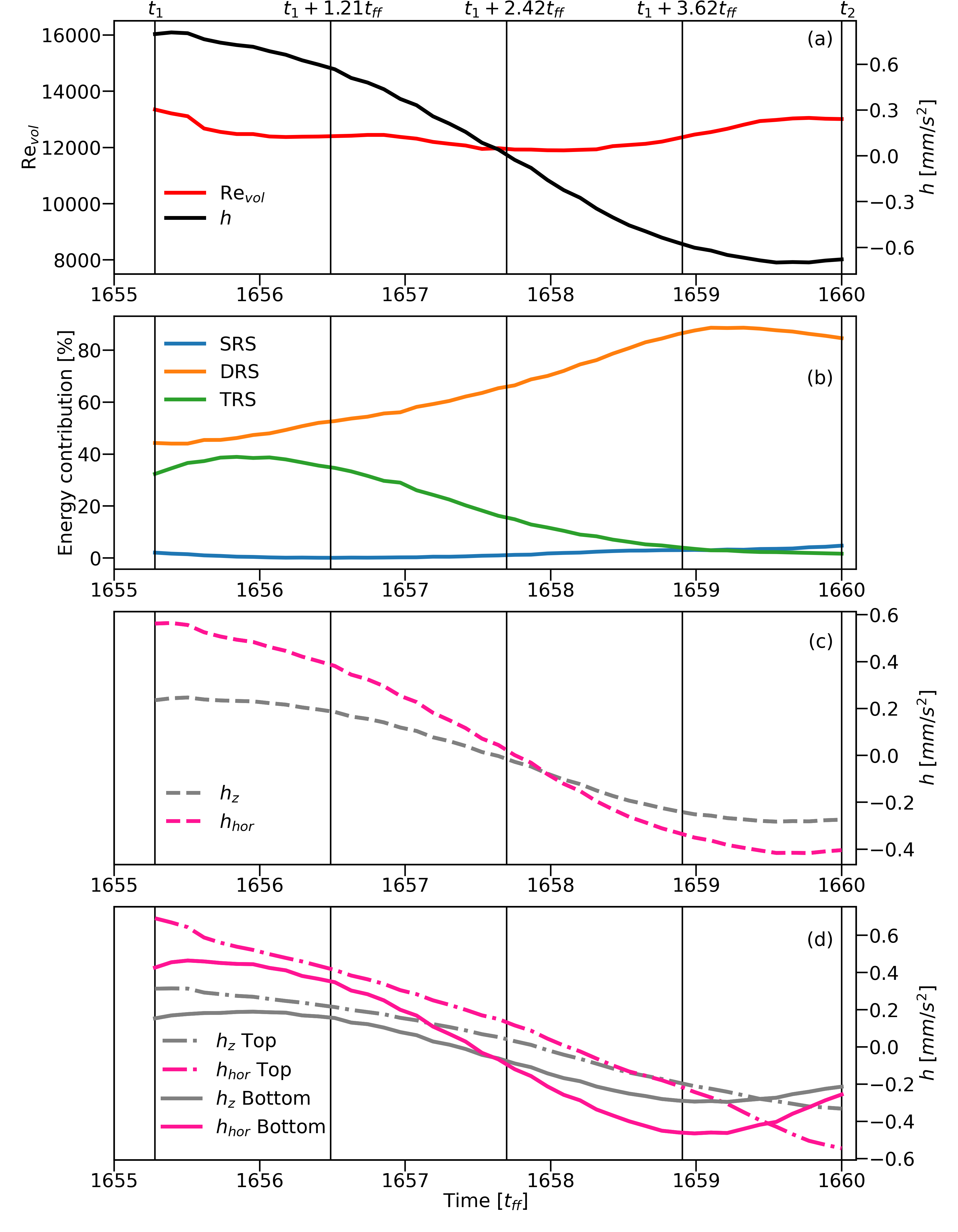}
    \centering
    \caption{(a) Global Reynolds number (red) and mean helicity density (black); 
    (b) Energy contributions of SRS (blue), DRS (orange), TRS (green); 
    (c) Vertical (dashed grey) and horizontal components of the mean helicity density (dashed pink); 
    (d) Vertical (grey) and horizontal (pink) components of the mean helicity density at 
    top half of the cylinder (dash-dotted) and bottom half of the cylinder (solid), 
    within the time interval $t_1$ to $t_2$ 
    of the experimental flow.}
    \label{fig:ra8_hr_hz_t1_t2}
\end{figure}

\begin{figure}[h!]
    \includegraphics[width=1\linewidth]{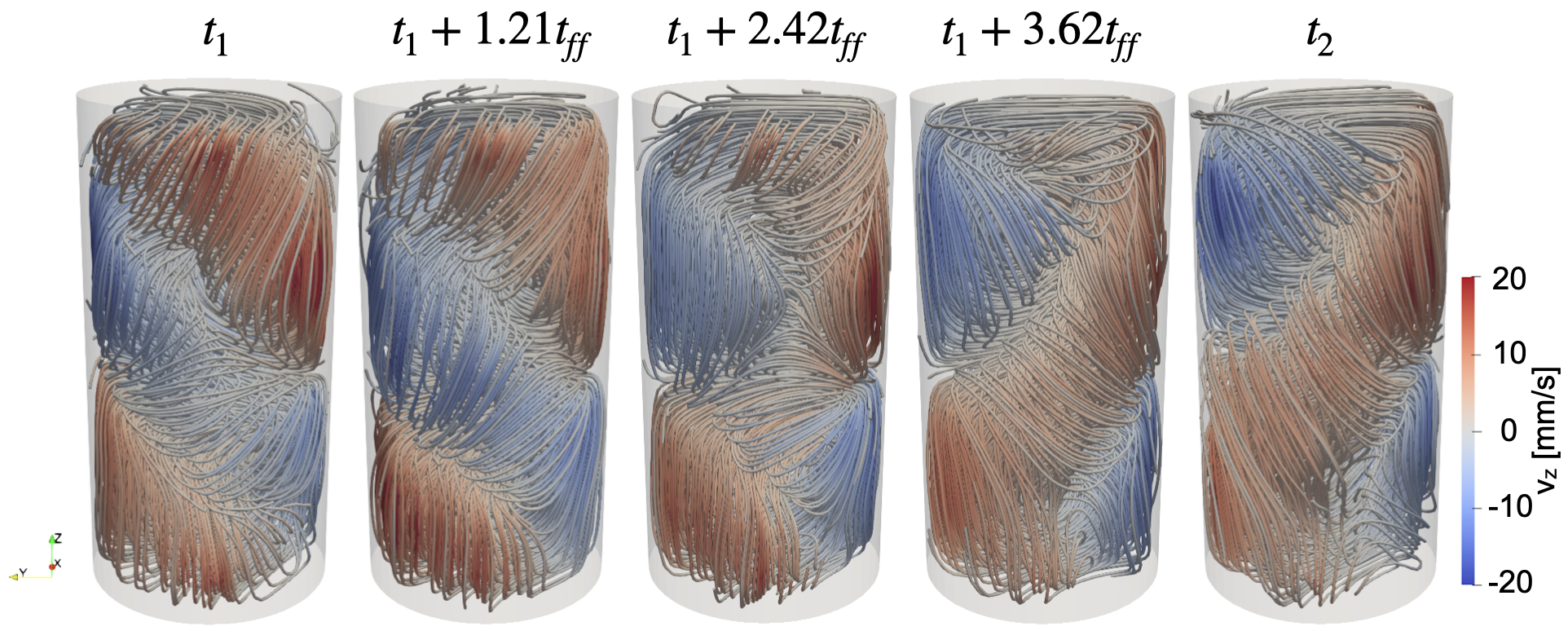}
    \centering
    \caption{Illustration of the helicity reversal at nearly constant Reynolds number for 
    a DRS, within the time interval $t_1$ to $t_2$ of the experimental flow. The snapshots correspond 
    to the five instants indicated by vertical lines in Fig. \ref{fig:ra8_hr_hz_t1_t2}.}
    \label{fig:reko_t1_t2}
\end{figure}

\begin{figure}[h!]
    \includegraphics[width=1\linewidth]{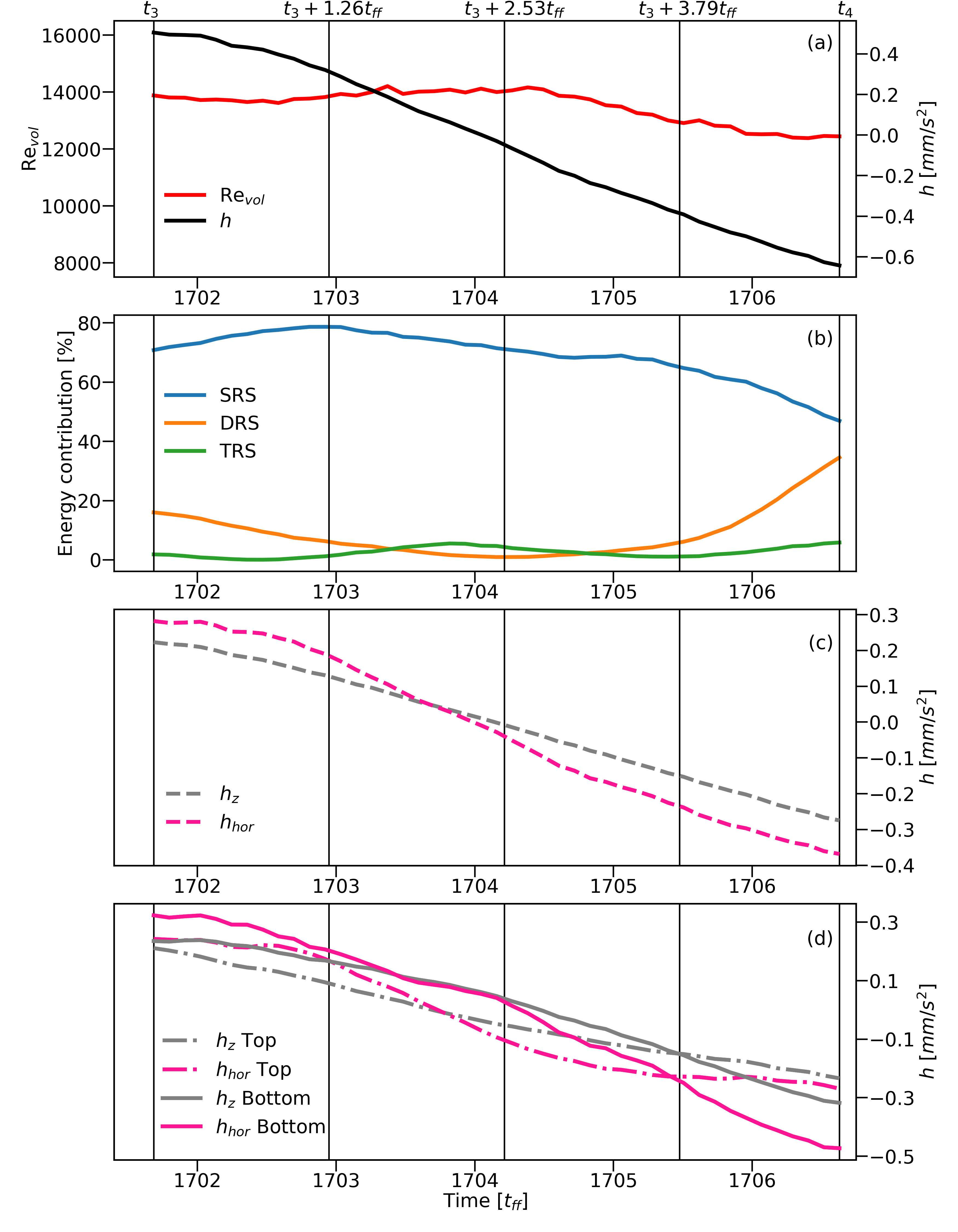}
    \centering
   \caption{(a) Global Reynolds number (red) and mean helicity density (black); 
    (b) Energy contributions of SRS (blue), DRS (orange), TRS (green); 
    (c) Vertical (dashed grey) and horizontal components of the mean helicity density (dashed pink); 
    (d) Vertical (grey) and horizontal (pink) components of the mean helicity density at 
    top half of the cylinder (dash-dotted) and bottom half of the cylinder (solid), 
    within the time interval $t_3$ to $t_4$ 
    of the experimental flow.}
    \label{fig:ra8_hr_hz_t3_t4}
\end{figure}

\begin{figure}[h!]
    \includegraphics[width=1\linewidth]{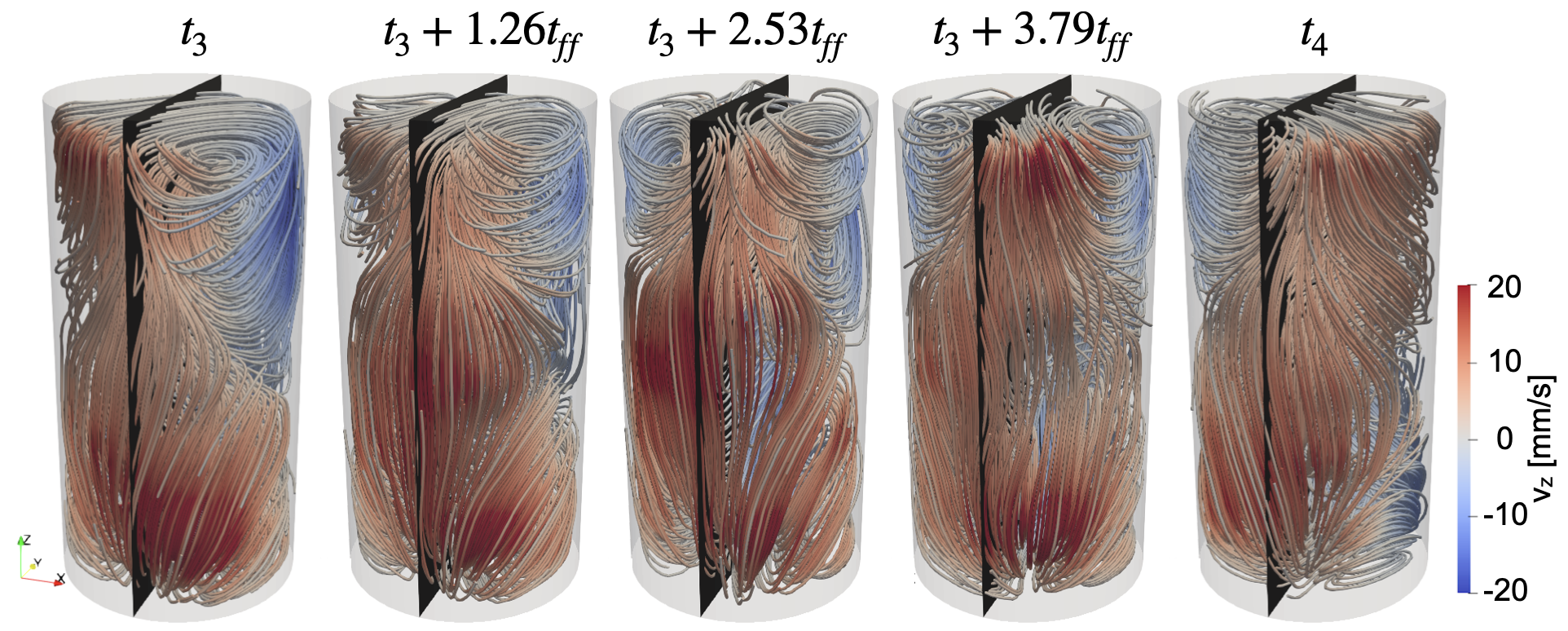}
    \centering
    \caption{Illustration of the helicity reversal at nearly constant Reynolds number for 
    a SRS, within the time interval $t_3$ to $t_4$ of the experimental flow. The snapshots correspond 
    to the five instants indicated by vertical lines in Fig. \ref{fig:ra8_hr_hz_t3_t4}.}
    \label{fig:reko_t3_t4}
\end{figure}

Helicity reversals in case of dominating DRS and  SRS 
were also observed in the experiment.
The corresponding time intervals $t_1$ to $t_2$ (Fig. \ref{fig:ra8_hr_hz_t1_t2}) and 
$t_3$ to $t_4$ (Fig. \ref{fig:ra8_hr_hz_t3_t4}) have been marked by black arrows 
in Fig. \ref{fig:h_re_modes}.
The time interval between $t_2$ and $t_1$ is approximately \num{4.8} $t_{\mathit{ff}}$, 
whereas the interval between $t_4$ and $t_3$ is around \num{5.1} $t_{\mathit{ff}}$.
Figs. \ref{fig:reko_t1_t2} and \ref{fig:reko_t3_t4} visualize the corresponding 
3D velocity 
fields for the two specified time intervals.   

Quite generally, the experimentally observed features are very similar to those
of the simulated flow. 
For the DRS, the horizontal component of the mean helicity density,
$h_{hor}$, is significantly stronger than the vertical one, $h_{z}$ 
(Fig. \ref{fig:ra8_hr_hz_t1_t2}(c)),
while their respective evolutions in the top and bottom half spaces
(Fig. \ref{fig:ra8_hr_hz_t1_t2}(d)) are quite parallel.
The change of twist in Fig. \ref{fig:reko_t1_t2} 
is more evident than in the case of 
simulations (Fig. \ref{fig:sim_t1_t2}),
which might have to do with 
the "contamination" of the DRS by SRS at the end
of the considered interval as discussed above.

As for the SRS, Fig. \ref{fig:ra8_hr_hz_t3_t4}(c)
shows again that $h_{hor}$ and  $h_{z}$ evolve very similarly, with only a slight preponderance
of $h_{hor}$. 
Also the evolutions in the top and bottom halves proceed
in parallel (Fig. \ref{fig:ra8_hr_hz_t3_t4}(d)), quite consistently with 
the simulation results from Fig. \ref{fig:sim_hr_hz_t3_t4}(d).
%
Again, the sloshing motion illustrated in the five snapshots
of Fig. \ref{fig:reko_t3_t4} appears somewhat smoother than the corresponding simulation results 
of Fig. \ref{fig:sim_t3_t4}.
%
\section{Conclusion}
In this study, we have analyzed the flow evolution 
in a slender RB convection cell which is characterized by frequent and
chaotic transitions between SRS, DRS and TRS.
Using both numerical simulations and experimental results, 
our main focus was on the relation between those specific flow modes 
and the helicity of the flow. 
Apart from concurrent changes
of flow modes and helicity, we also observed helicity
reversals within a given flow mode when the Reynolds number 
is nearly constant. 

For the SRS, the helicity reversal goes along with a clearly visible
sloshing motion of the elongated "flywheel". 
Yet, 
in addition to its corresponding horizontal part, the reversing 
helicity contains also an equally strong vertical part,
pointing to a torsional component of the flow.
A full period of these changes 
(i.e., the double of the observed reversal times) 
is in the range of \num{8} free-fall times,
in good agrement with previous results of Wondrak et al. 
\cite{Wondrak2023}
In any case, the helicity parts in the
top and bottom half of the cylinder behave similarly.

For the DRS, the helicity reversals, which correspond to a simultaneous
twisting of the two flow cells which are stacked over each other,
proceed a bit faster than in case of the SRS.
The horizontal contribution of the helicity is typically stronger than the 
vertical one. 
As in the SRS case, the evolutions in the top
and bottom halves are in phase.
Remarkably, the flow and helicity evolution in this DRS case is 
very similar to that in case of the kink-type, current-driven 
TI, as previously observed by Weber et al. \cite{Weber2015}
(see, in particular, the animated version of Fig. 13,
available in the supplementary material of that paper).
Hence, whatever the specific driving agent of the DRS might ever be,
its helicity evolution turns out to be a quite generic and 
universal feature.

Investigations of potential tidal synchronization 
for the slender RB cell, as 
previously carried out for the 
${\Gamma}$ = \num{1} cell \cite{Juestel2022}, 
are planned for the future.
The same applies to the corresponding 
helicity evolution for flows with 
${\Gamma}$ $\gg$ \num{1}  which might indeed be 
of high relevance for the so-called 
small-scale dynamo in the shallow 
subsurface layer of the Sun \cite{Kitiashvili2015}.
%
\begin{acknowledgements}
The authors would like to express their gratitude to Thomas Gundrum and Stefanie Sonntag for their invaluable technical support throughout the course of the experiment. 
Additionally, R.M. would like to extend their appreciation to Ashish Mishra for engaging in insightful discussions that enriched the research. 
\end{acknowledgements}
\section*{Funding}
This work is supported by the Deutsche Forschungsgemeinschaft (DFG) under the grant VO 2331/1.
\section*{Conflict of Interest}
The authors have no conflicts to disclose.
\section*{Data Availability Statement}
The data that support the findings of this study are available from the corresponding author upon reasonable request.
\section*{Author Contributions}
\textbf{Rahul Mitra}: Conceptualization (support), Data curation (lead), 
Formal analysis (lead), Investigation (lead), Methodology (support), 
Software (equal), Visualization (lead), Writing - original draft (equal),
Writing - review \& editing (equal)
\textbf{Frank Stefani}: Conceptualization (lead), Methodology (lead), 
Project administration, Supervision (lead), Visualization (support), 
Writing - original draft (equal), Writing - review \& editing (equal)
\textbf{Vladimir Galindo}: Software (equal), Writing - review \& editing (equal)
\textbf{Sven Eckert}: Supervision (support), 
Writing - review \& editing (equal)
\textbf{Max Sieger}: Investigation (equal), Supervision (support), 
Writing - review \& editing (equal)
\textbf{Tobias Vogt}: Funding acquisition (lead), 
Supervision (support), Writing - review \& editing (equal)
\textbf{Thomas Wondrak}: Methodology (support), 
Supervision (support), Writing - review \& editing (equal)
All authors have read and agreed to the published version of the manuscript.
%

\appendix
\section{Fourier spectra of simulation data}\label{appendix:fourier_simulation}
\begin{figure}[b!]
    \includegraphics[width=1\linewidth]{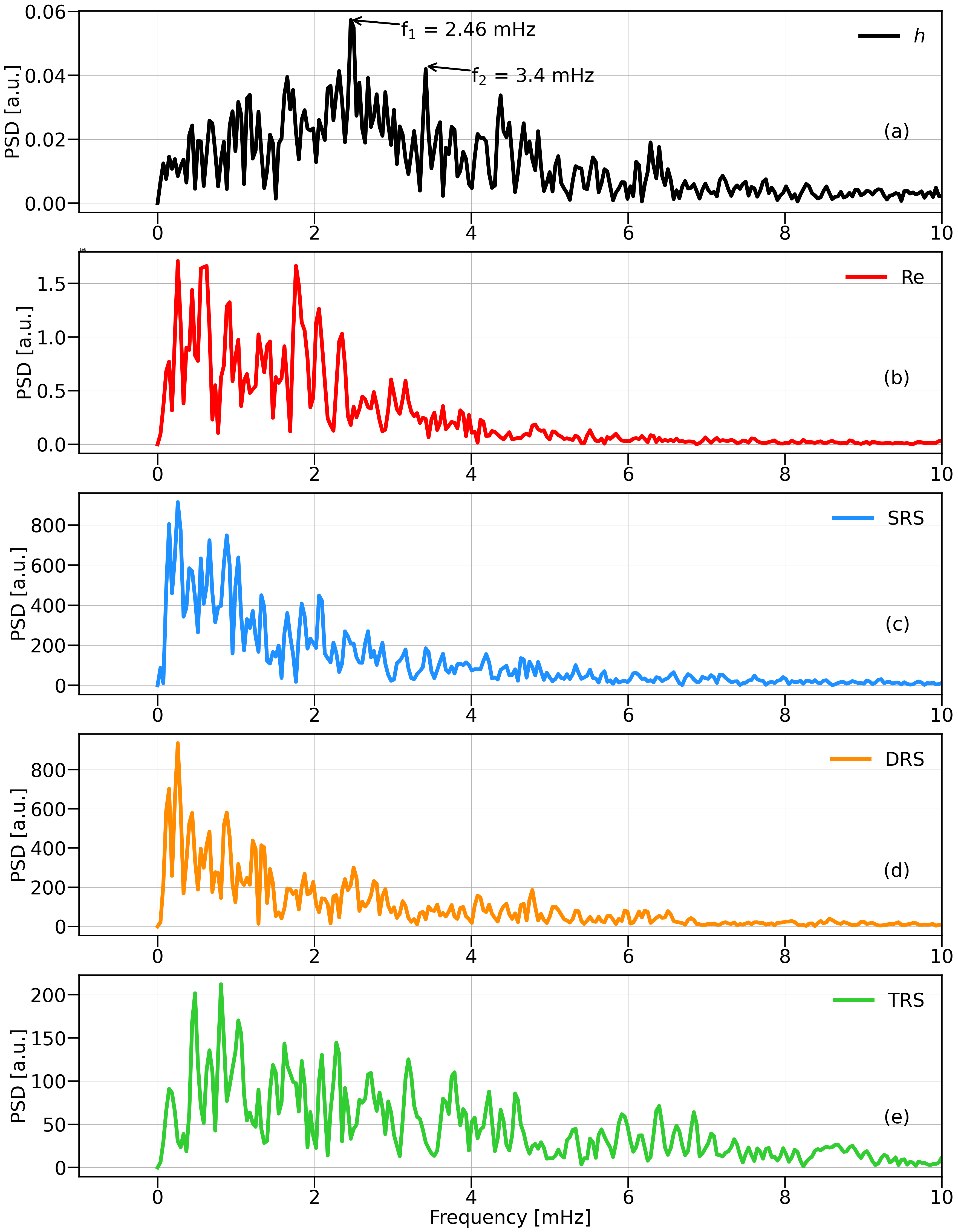}
    \centering
    \caption{Fourier spectra of (a) helicity (black), (b) Reynolds number (red), 
    energy contribution of (c) SRS (blue), (d) DRS (orange) and (e) TRS (green) 
    for the simulation.}
    \label{fig:fourier_sim}
\end{figure}
A Fourier spectral analysis was performed to discern the 
time-dependence of the helicity, the Reynolds number, and 
the energy contributions of single, double, and triple roll states 
within the turbulent flow. 
The Fourier spectra of the mean helicity density, Reynolds number and the energy 
contributions of the flow states are shown in Fig. \ref{fig:fourier_sim}.

The first observation concerns a very prominent peak at \SI{2.46}{\milli\hertz} 
(corresponding to nearly \num{8} $t_{\mathit{ff}}$) observed in 
the helicity spectrum which signifies a substantial periodic contribution to the 
helicity of the flow. 
Looking back on Figs. \ref{fig:sim_hr_hz_t3_t4} and \ref{fig:sim_t3_t4}
(where the period of half an oscillation was \num{4.6} $t_{\mathit{ff}}$)
it appears that this period is quite representative for 
a typical helicity oscillation within an SRS. 
Notably, such a period is also consistent with the work of Wondrak et al. \cite{Wondrak2023}
where a typical change rate of the LSC's angle of \ang{40}-\ang{50}/$t_{\mathit{ff}}$ had 
been observed (see page 23 of that paper).

By contrast, the observed period for the DRS case 
(see Figs. \ref{fig:sim_hr_hz_t3_t4} and \ref{fig:sim_t3_t4})
was \num{5.7} $t_{\mathit{ff}}$, corresponding to \SI{3.4}{\milli\hertz}.
A secondary peak close to this value is visible in Fig. \ref{fig:fourier_sim}(a).

At any rate, the typical peaks for the Reynolds 
number in Fig. \ref{fig:fourier_sim}(b) are at signficantly 
lower frequencies than those for the helicity. 
One of those peaks, at \SI{1.7}{\milli\hertz},
shows also up in the spectrum of the SRS energy (Fig. \ref{fig:fourier_sim}(c))
which indicates that the spectrum of the Reynolds number is dominated
by the transitions of different flow types. 
This is confirmed by 
a closer inspection of the typical transition times 
between the flow modes in Fig. \ref{fig:h_re_modes_sim}(b).
\section{Helicity comparison}\label{appendix:helicity_simulation_comparison}

To investigate the reliability of CIFT for reconstructing the helicity of the flow, 
we have used the simulated velocity field (see section \ref{sec:simulation}), 
from which we generated a synthetic CIFT reconstruction 
with the same excitation magnetic field and sensor arrangement 
as used in the experiment (see section \ref{sec:experiment:setup}). 
The mean helicity density of the simulated flow can then 
be compared to that of the flow reconstructed by CIFT.
\begin{figure}[h!]
    \includegraphics[width=1\linewidth]{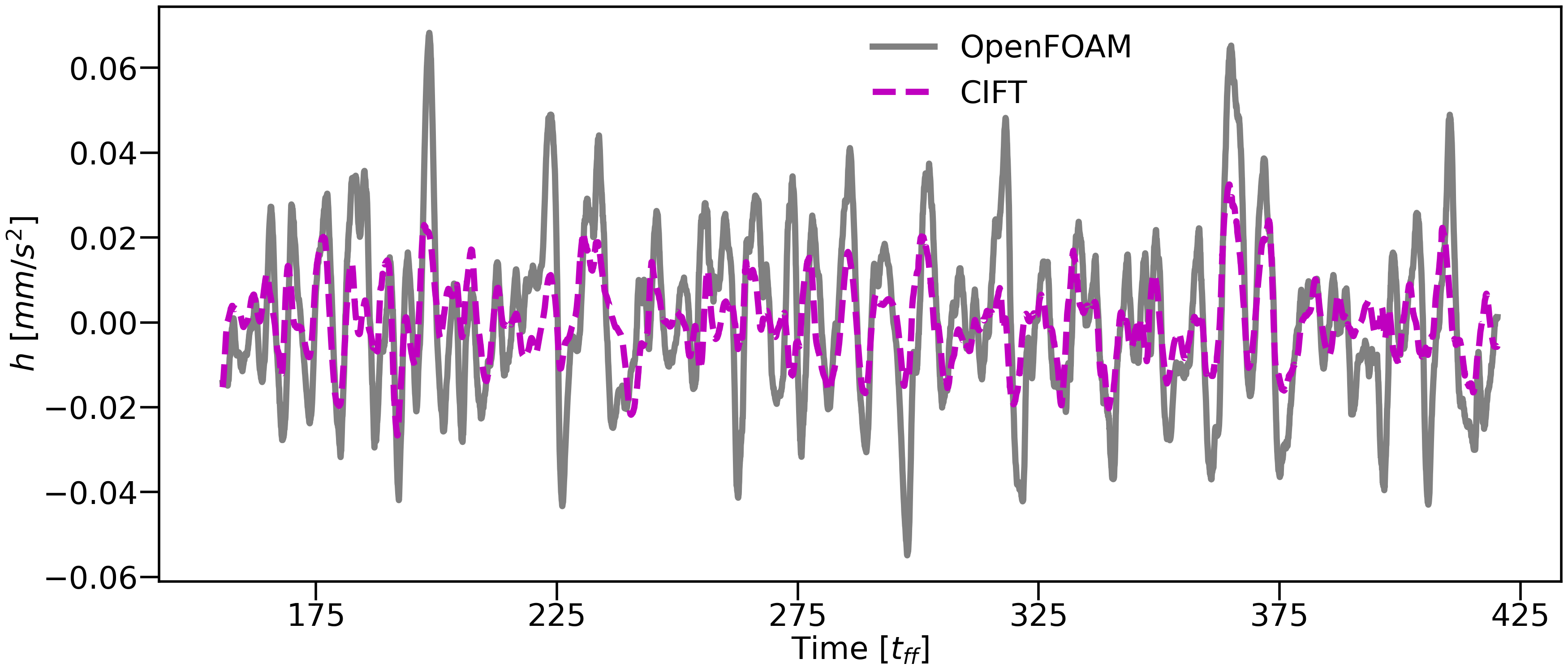}
    \centering
    \caption{Time-dependent mean helicity density of the simulated velocity 
    field (solid, grey) and of the 
    corresponding CIFT-reconstructed field (dashed, magenta).}
    \label{fig:helicity_comparison}
\end{figure}

Fig. \ref{fig:helicity_comparison} illustrates the time variation of the 
mean helicity density of the simulated (solid, grey) and the reconstructed flow (dashed, magenta), 
over the \SI{13600}{\second} of chosen simulation time.
The general variations are appreciably reconstructed by CIFT.
The mean correlation is \num{66.4}\%, which is in accordance with the 
mean quality of reconstruction for this particular sensor configuration 
and excitation scheme \cite{Mitra2024}.
%

\bibliography{references}

\end{document}